\documentclass[final]{siamltex}

\usepackage{amsmath, amssymb, dsfont, bm}
\usepackage{graphicx}
\usepackage{color}

\newcommand {\R} {{\mathds R}}

\newcommand {\C} {{\mathds C}}

\newcommand {\BE} {\begin{eqnarray}}
\newcommand {\EE} {\end{eqnarray}}

\newcommand {\BS} {\begin{eqnarray}}
\newcommand {\ES} {\end{eqnarray}}

\newcommand {\BEW} {\begin{eqnarray*}}
\newcommand {\EEW} {\end{eqnarray*}}

\newcommand {\x} {{\bm x}}

\newcommand {\TF} {{{TF}}}

\renewcommand {\phi} {{\varphi}}

\newcommand{\pl}{\partial }

\title{A new Sobolev gradient method for direct minimization of the Gross--Pitaevskii energy with rotation}

\author{Ionut Danaila and Parimah Kazemi\thanks{UPMC Univ Paris 06, UMR 7598, Laboratoire Jacques-Louis Lions, F-75005, Paris, France and
              CNRS, UMR 7598, Laboratoire Jacques-Louis Lions, F-75005, Paris, France
              }}

\begin{document}

\maketitle

\begin{abstract}
In this paper we improve  traditional steepest descent methods for the direct minimization of the Gross-Pitaevskii (GP) energy with rotation at two levels. We first define a new inner product to equip the Sobolev space $H^1$ and derive the corresponding gradient. Secondly, for the treatment of the mass conservation constraint, we use a projection method that avoids more complicated approaches based on modified energy functionals or traditional normalization methods. The descent method with these two new ingredients is studied theoretically in a Hilbert space setting and we give a proof of the global existence and convergence in the asymptotic limit to a minimizer of the GP energy.
The new method is implemented in both finite difference and finite element two-dimensional settings and used to compute various complex configurations with vortices of rotating Bose-Einstein condensates. The new Sobolev gradient method shows better numerical performances compared to classical $L^2$ or $H^1$ gradient methods, especially when high rotation rates are considered.
\end{abstract}


%

\pagestyle{myheadings}
\thispagestyle{plain}
\markboth{I. DANAILA AND P. KAZEMI}{A NEW SOBOLEV GRADIENT METHOD}


\section{Introduction}

First  experimental realizations of Bose-Einstein condensates (BECs) in dilute alkali-metal gases  \cite{anderson,bradley,davis} led to an explosion of mathematical and theoretical studies aimed at better understanding such systems.  Recent efforts were devoted to documenting the superfluid nature of the condensate by providing evidence for the existence of quantized vortices when rotating the condensate. It was indeed experimentally observed \cite{madison,matthews,aboshaerr,haljan} that instead of solid body rotation, the condensate rotates by forming vortices with quantized circulation.
Initially, a few vortices are formed, and, as the rotation frequency increases, the vortices form an array similar to the Abrikosov lattice observed in type II superconductors. Since the rotating BEC is a highly controllable system with a simple theoretical description, it provides a perfect set-up for the theoretical study of macroscopic systems with quantized vortices.

In the zero-temperature limit, a  dilute gaseous BEC is mathematically described by a macroscopic wave function  derived in the framework of the  Gross--Pitaevskii (GP) mean field theory. The spatial configuration of the wave function $\psi(\mathbf{x})$, with $\mathbf{x}=(x, y, z)^t$, is obtained by minimizing the GP energy in the rotating frame,
\begin{equation}
E(\psi) = \int_{\R^3} \frac{\hbar^2}{2m} | \nabla \psi |^2 + \tilde{V}_{trap}|\psi|^2 + \frac{\tilde{g}}{2}|\psi|^4 +  i \hbar \psi^* \tilde{\mathbf{\Omega}}\cdot(\mathbf{x} \times \nabla)\psi,
\end{equation}
subject to the normalization condition, $\int_{\R^3} | \psi|^2 = N$, with $N$ the number of particles (atoms).   In the previous expression, $\hbar$ is Planck's constant, $m$ the atomic mass of the gas, $\tilde{\mathbf{\Omega}}$ the angular velocity vector, and $\tilde{V}_{trap}$ the magnetic trapping potential with trap frequencies ($\omega_x, \omega_y, \omega_z$).  We denote by $\psi^*$ the complex conjugate of $\psi$. The interactions between atoms are described by $\tilde{g}=\frac{4 \pi \hbar^2 a_s}{m}$, with $a_s$ the $s$-wave scattering length. As in most of experimental settings, we consider in the following that $\tilde{\mathbf{\Omega}} = (0, 0, \tilde{\Omega})^t$ and that $\tilde{V}_{trap}$ has a lower bound and $\tilde{V}_{trap}(\mathbf{x}) \rightarrow \infty$, as $\mathbf{x} \rightarrow \infty$. Since from the previous assumption we can infer that  $\psi(\mathbf{x}) \rightarrow 0$, as $\mathbf{x} \rightarrow \infty$, it suffices to work in a bounded domain ${\cal D} \subset \R^3$ with homogeneous Dirichlet boundary conditions $\psi=0$ on $\partial D$.

In practice it is common to scale the energy so that the units become dimensionless. Using the scaling  $\mathbf{r} = \mathbf{x}/d$, $u(\mathbf{r})=\psi(\mathbf{x}) d^{3/2}/\sqrt{N}$, $\Omega = \tilde{\Omega}/\omega_\perp$, with $d=\sqrt{\frac{\hbar}{m \omega_{\bot}}}$  the harmonic-oscillator length and $\omega_{\perp}=\min(\omega_x, \omega_y)$ the transverse trap frequency, we obtain the non-dimensional energy (per particle) functional:
\begin{equation}
E(u) = \int_{\cal D} \frac{| \nabla u |^2}{2} + V_{trap}|u|^2 + \frac{g}{2}|u|^4 -   \Omega i u^* (A^t \nabla) u,
\label{gp}
\end{equation}
where $V_{trap}=\frac{1}{\hbar \omega_{\bot}} \tilde{V}_{trap}$, $g=\frac{4 \pi N a_s}{d}$, and  $A=(y,-x, 0)$.
The mass conservation constraint becomes:
\begin{equation}
\int_{\cal D} |u|^2 = \| u \|^2 = 1,
\label{eq-psi-norm}
\end{equation}
where we denote by $\| . \| = \| . \|_{L^2({\cal D}, \C)}$.

For given constants  $\Omega, g$, and  trapping potential function $V_{trap}$, the minimizer $u_g$ of the functional (\ref{gp}) under the constraint (\ref{eq-psi-norm}) is called the ground state of the condensate. Local minima of the energy functional with energies larger that $E(u_g)$ are called excited (or metastable)  states of the condensate. For a detailed discussion of the derivation of the Gross-Pitaevskii energy and the physics of rotating Bose-Einstein condensates see, for example, \cite{fetter} and \cite{lieb}.

The two key issues in numerically computing ground or excited states of BEC are {\em (i)} how to derive a numerical algorithm that starts from a chosen initial state and iteratively diminishes the energy of the solution to rapidly converge to a local minimum of the functional (\ref{gp}), and {\em (ii)} how to take into account the mass constraint (\ref{eq-psi-norm}).
These two issues are obviously connected and have to be considered together in deriving efficient numerical algorithms.  We present in this paper new approaches to address both issues and prove their superior numerical performance in the case of the energy minimization of the Gross-Pitaevskii energy with rotation.

Most of the numerical algorithms proposed in the literature use
the so-called {\em normalized gradient flow} \cite{bao}, that consists in two steps:
the steepest descent method is applied to the unconstrained problem,
\begin{equation}
\frac{\partial u}{\partial t} = - \frac{1}{2} \frac{\partial E(u)}{\partial u} = \frac{\nabla^2 u}{2} - V_{trap} u - g  |u|^2u  +   i \Omega A^t \nabla u,
\label{eq-grad-flow}
\end{equation}
to advance the solution from the discrete time level $t_n$ to $t_{n+1}$; the obtained predictor ${\tilde u}(r,t_{n+1})$ is then normalized
in order to satisfy the unitary norm constraint and set the solution at $t_{n+1}$:
\begin{equation}
u(\mathbf{r},t_{n+1}) \triangleq \frac{{\tilde u}(\mathbf{r},t_{n+1})}{\|{\tilde u}(\mathbf{r},t_{n+1})\|}.
\label{eq-steep-norm}
\end{equation}

The gradient flow equation (\ref{eq-grad-flow}) (or the related {\em continuous gradient flow} equation, see \cite{bao}) can be viewed as a complex heat equation and, consequently, solved by different classical time integration schemes (Runge-Kutta-Fehlberg \cite{garcia-ripoll}, backward Euler \cite{bao,du,bao2008}, second-order Strang time-splitting  \cite{bao,du}, combined Runge-Kutta-Crank-Nicolson scheme \cite{danaila1,danaila2,danaila3}, etc.), and different spatial discretization methods (Fourier spectral \cite{garcia-ripoll}, finite elements \cite{du}, finite differences \cite{bao,danaila1,danaila2,danaila3}, sine-spectral \cite{bao}, Laguerre--Hermite pseudo-spectral \cite{bao2008}, etc.).

It is interesting to note that in the descent method (\ref{eq-grad-flow}), the right-hand side represents the $L^2$-gradient (or {\em ordinary} gradient) of the energy functional. An important improvement of the convergence rate of the descent method was obtained in \cite{garcia-ripoll} by replacing  the ordinary gradient with the gradient defined on the Sobolev space $H^1({\cal D}, \C)$. The same Sobolev gradient method (see \cite{jwn} for various applications of this method) was recently used to minimize simpler Schr{\"o}dinger type functionals in \cite{lookman}. Similar increase of the convergence rate over the ordinary gradient method was reported. A first new contribution of the present paper is to introduce  a new definition of the inner product to equip the Sobolev space $H^1$ in the case of the GP energy with rotation. A proof of the existence of the asymptotic limit for the evolution equations associated with the Sobolev gradients in a Hilbert space setting is also given. When implemented in a finite difference or finite element settings, the new Sobolev gradient method shows better numerical performances
compared to classical $L^2$ or $H^1$ gradient methods, especially when high rotation rates ($\Omega$) are considered.

The second important contribution of this work concerns the issue of the mass conservation constraint (\ref{eq-psi-norm}). Instead of the classical (and very popular) normalization approach (\ref{eq-steep-norm}), we suggest a projection method that preserves the norm of the initial state through the minimization procedure.
The idea to project the (Sobolev) gradient into the {\em tangent space} associated to the constraint was already used to derive numerical algorithms for minimizing harmonic maps \cite{alouges1,pierre}, and, recently, to numerically find the smallest eigenvalue and corresponding eigenvectors of a Hermitian operator \cite{alouges}. Different algorithms based on the projected gradient were developed in these studies and successfully applied to different energy functionals: the Oseen-Frank energy for liquid crystals \cite{alouges1}, the Dirichlet energy of harmonic maps \cite{pierre} and the Hartree-Fock energy for quantum chemical molecular systems \cite{alouges}.
We derive here a projected Sobolev gradient method adapted to the Gross-Pitaevskii energy functional and provide an
explicit expression of the projected gradient that allows to minimize trajectories when  Hilbert spaces other than $L^2$ are considered.
The new projection method proved very helpful in numerical implementations and allowed to avoid alternative methods to treat the mass constraint by adding to the energy functional a penalty term with a Lagrange multiplier (e. g. \cite{garcia-ripoll,bao1}).

The organization of the paper is as follows.  In section \ref{sec-gradients} we introduce an alternate inner product on the Sobolev space $H^1$ and show that this inner product is equivalent to the traditional inner product on $H^1$.  The corresponding new Sobolev gradient is also derived.  We  discuss in section \ref{sec-minim} a constructive projection method  for  the mass constraint and give our existence and convergence result for the asymptotic limit of the evolution equation defined by the Sobolev gradients.  In section \ref{sec-implement} we give a discussion of the finite difference  and finite element implementations in two-dimensions.  The last section is devoted to numerical tests designed as benchmarks to compare performances of different Sobolev gradient methods. The effectiveness of the newly proposed Sobolev gradient method is proved by computing stationary states of rotating BEC that are physically relevant (high rotation and large interaction constants).

\section{Gradient descent methods using several gradients}\label{sec-gradients}

In optimization problems that use a gradient descent or ascent technique, one usually has a choice of norms to use in the argument.  If the norm  has an associated inner product, then one can obtain a gradient with respect to this inner product (see \cite{jwn} for an explanation).  In the example of the minimization problem of Schr{\"o}dinger type functionals, the gradient represents the direction of change per unit time.  Therefore, one wants to choose the gradient in the descent method \eqref{eq-grad-flow} so that the change in energy is maximal at each step.  For the case of the Gross-Pitaevskii energy with rotation, we notice that the energy \eqref{gp} can be written as
\begin{equation} \label{gpcov}
E(u)= \int_{\cal D} \frac{| \nabla u + i \Omega A^t u|^2}{2} + V_{eff}|u|^2 +\frac{g}{2}|u|^4
\end{equation}
where, the {\em effective} trapping potential is defined as:
\begin{equation}
V_{eff}(r)= V_{trap}(r) - \frac{ \Omega^2 r^2}{2}.
\end{equation}
This form of the energy suggests the definition of
a new norm  to equip the domain of the functional such that the functional is coercive with respect to this norm.  This implies that if the size of the argument is large, then naturally the value of the functional will be large as well, making it suitable for rotating cases.

\subsection{Inner products  and norms}\label{subsec-norm}

We define three inner products on $C^1({\cal D},\C)$ and study the completion of this space with respect to the norm arising from each of these inner products.  Consider the  inner products:
\begin{equation}
\langle u , v \rangle_{L^2} = \int_{\cal D} \langle u , v \rangle,
\label{eq-innpL}
\end{equation}
\begin{equation}
\langle u , v \rangle_{H} = \int_{\cal D} \langle u, v \rangle + \langle \nabla u , \nabla v \rangle,
\label{eq-innpH}
\end{equation}
and
\begin{equation}
\langle u , v \rangle_{H_A} = \int_{\cal D} \langle u , v \rangle  + \langle \nabla_A u , \nabla_A v \rangle,
\label{eq-innpHA}
\end{equation}
where $\nabla_A = \nabla + i \Omega A^t$, $\Omega$ is a fixed positive number. Here $\langle \cdot , \cdot \rangle$ denotes the complex inner product.  Each of these inner products leads to a norm which we will denote by $\| \cdot \|_{L^2},  \| \cdot \|_{H},$ and $\| \cdot \|_{H_A}$.  For $X=L^2, H, H_A$, consider the completion of $\{ u \in C^1({\cal D}, \C): \|u\|_X < \infty \}$ with respect to each of the respective norms.  In the first case, one obtains the Hilbert space $L^2=L^2({\cal D}, \C)$, in the second case $H^1=H^{1,2}({\cal D}, \C)$, and in the third case we call the resulting Hilbert space $H_A=H_A({\cal D}, \C)$ (see \cite{adams} for details on Sobolev spaces).  Furthermore, the following calculation shows how the three norms are related.  We first note that:
\begin{equation}
\langle \nabla_A u , \nabla_A v \rangle =
\langle \nabla u , \nabla v \rangle + \Omega ^2 r^2\langle u , v \rangle +
i \Omega (\langle A^tu, \nabla v \rangle - \langle \nabla u , A^t v \rangle).
\label{gradAuv}
\end{equation}
If $r_{\cal D}$ denotes the radius of $D$,  one has
\begin{equation}
\langle \nabla_A u , \nabla_A u \rangle =|\nabla u + i \Omega A^t u|^2 \leq 2( |\nabla u |^2 + r_{\cal D}^2 \Omega^2 |u|^2),
\end{equation}
and, consequently,
\begin{equation}
\| u \|_{H_A}  \leq \int_{\cal D} (1+2 r_{\cal D}^2 \Omega^2)|u|^2 + 2 |\nabla u|^2 \leq c \|u\|_H^2
\end{equation}
where $c=max(1+2 r_{\cal D}^2 \Omega^2, 2)$.  Hence one has that the $H^1$ norm dominates the $H_A$ norm.

In the same time, using the identity
\begin{equation} \label{identity}
\int_{\cal D} \langle A^t u ,\nabla v \rangle  = - \int_{\cal D} \langle  \nabla u , A^t v \rangle,
\end{equation}
we infer from \eqref{gradAuv} that
\begin{equation} \label{HAidentity}
\int_{\cal D} \langle \nabla_A u , \nabla_A v \rangle =
\int_{\cal D} \langle \nabla u , \nabla v \rangle + \int_{\cal D} \Omega ^2 r^2\langle u , v \rangle -
2i \Omega \int_{\cal D} \langle \nabla u, A^t  v \rangle.
\end{equation}
and, consequently,
\begin{equation}
\int_{\cal D} |\nabla u + i \Omega A^t u|^2 =
\int_{\cal D} |\nabla u|^2  + \Omega^2 r^2 |u|^2 - 2 \Omega \langle i \nabla u , A^t  u \rangle.
\end{equation}
Also, for $\epsilon > 0$, one has the inequality
\begin{equation}
ab = \frac{a}{\epsilon} b \epsilon \leq \frac{1}{2}\left( \left(\frac{a}{\epsilon}\right)^2 + (b \epsilon)^2\right).
\end{equation}
Now, using the above inequality and Cauchy-Schwartz one has that
\begin{equation}
2|\langle i \nabla u , A^t u \rangle| \leq 2|\nabla u| |A^t u| \leq (\epsilon|\nabla u|)^2 + \frac{| A^t u|^2}{\epsilon^2}.
\end{equation}
Thus
\begin{equation}
- 2 \Omega \langle i \nabla u , A^t  u \rangle \geq - \Omega((\epsilon|\nabla u|)^2 + \frac{| A^t u|^2}{\epsilon^2}) \geq   -\Omega((\epsilon|\nabla u|)^2 + \frac{r_{\cal D}^2}{\epsilon^2}|u|^2).
\end{equation}
From this one has that
\begin{eqnarray}
\int_{\cal D} |\nabla u + i \Omega A^t u|^2 \geq \int_{\cal D} |\nabla u|^2 + \Omega^2 r^2 |u|^2 - \Omega((\epsilon|\nabla u|)^2 + \frac{r_{\cal D}^2}{\epsilon^2}|u|^2) = \\
\int_{\cal D} (1-\Omega \epsilon^2)|\nabla u|^2 +  (\Omega^2r^2 - \Omega \frac{r_{\cal D}^2}{\epsilon^2})|u|^2 \geq \\
\int_{\cal D} (1-\Omega \epsilon^2)|\nabla u|^2 -   \Omega \frac{r_{\cal D}^2}{\epsilon^2}|u|^2
\end{eqnarray}
Now we choose $\epsilon$ so that $0 < 1 - \Omega \epsilon^2 < 1$ and let $k = 1+  \frac{\Omega}{\epsilon^2}r_{\cal D}^2 $. Since $k > 1$, we can write
\begin{eqnarray}\nonumber
k \int_{\cal D} |u|^2 + | \nabla_A u|^2 > \int_{\cal D} k|u|^2 + |\nabla_A u|^2 \geq\\
 \int_{\cal D} |u|^2 + (1-\Omega \epsilon^2)| \nabla u|^2 > (1-\Omega \epsilon^2)\int_{\cal D} |u|^2 + |\nabla u|^2.
\end{eqnarray}
and infer that the $H_A$ norm dominates the $H^1$ norm.  Hence the two norms are equivalent.  Furthermore, we have the following relationship between the three Hilbert spaces,
\begin{equation}
H^{1,2}({\cal D} , \C) = H_A({\cal D},\C) \subset L^2({\cal D},\C).
\end{equation}
As sets $H^1$ and $H_A$ are equal.  However, by using the equivalent norm induced on $H_A$, we will see that the numerical performance of the descent method is improved for the minimization of the GP energy with rotation.

\subsection{Gradients}\label{subsec-grad}

The next step in writing a descent method to directly minimize the energy, as given in equation \eqref{gp} or \eqref{gpcov}, is to obtain a gradient corresponding to each inner product. Taking the Fr{\'e}chet derivative of \eqref{gp}, one gets that
\begin{equation}\label{derivgp}
E'(u)h= \int_{\cal D} \Re \ (\langle  \nabla u , \nabla h \rangle + \langle  2V_{trap} \ u + 2g  |u|^2u  -   2i \Omega A \nabla u , h\rangle)
\end{equation}
or equivalently
\begin{equation} \label{derivgpcov}
E'(u)h= \int_{\cal D} \Re \ ( \langle \nabla_A u ,  \nabla_A h \rangle  + \langle 2V_{eff} \ u + 2g |u|^2u, h \rangle).
\end{equation}
Since $E'(u)$ is a continuous linear functional from $H^1$ to $\R$ then for each $u \in H^1$, there exists a unique member of $H^1$ which we denote by $\nabla_H E(u)$ so that
\begin{equation}
E'(u)h = \Re \langle  \nabla_H E(u), h \rangle_H,
\end{equation}
for all $h \in H^1$.  We say that $\nabla_H E : H^1 \rightarrow H^1$ is a gradient for $E$ taken with respect to the $H^1$ inner product. Likewise $E'(u)$ is a continuous linear functional from $H_A$ to $\R$, thus it has a representation like the one given above.  We denote this gradient by $\nabla_{H_A} E : H_A \rightarrow H_A$ (see \cite{jwn} for a background on gradients obtained in this manner).  Furthermore, we note from \eqref{derivgp} that for all $h \in C^{\infty}_c({\cal D}, \C)$ one has that
\begin{equation} \label{el}
E'(u)h= \Re \langle \nabla_X E, h \rangle_{X}= \int_{\cal D} \ \Re \langle  -\nabla^2 u + 2V_{trap} u + 2g  |u|^2u  -   2 i \Omega A^t \nabla u, h \rangle.
\end{equation}
When $X=L^2$, we directly obtain the expression of $\nabla_{L^2} E$, the $L^2$ (or {\em ordinary}) gradient of $E$, already recalled in \eqref{eq-grad-flow}. From a practical point of view, it is interesting to note that $H^1$ and $H_A$ gradients will be computed using different forms of \eqref{el}: the corresponding strong formulation for the finite difference implementation (see also \cite{garcia-ripoll}) and the weak formulation for the finite element implementation (see also \cite{lookman}).

\section{Constrained energy minimization}\label{sec-minim}

\subsection{Projection method for the mass constraint}\label{subsec-mass}

Before discussing the gradient descent method, we give a brief description of the projection used to deal with the mass constraint.  In approximating stationary states, one could in principle use a normalized gradient flow in conjunction with a traditional Lagrange multiplier for the constraint \cite{garcia-ripoll,bao1}. We adopt here a different approach and develop a projection method that will, in the continuous case,  enforce the constraint for all time.

The method for enforcing the constraint is presented in \cite{jwn} for any general constraint and hence does not provide the needed expression for our case. For the unitary norm constraint, several projected gradient methods are developed in \cite{pierre,alouges}, based on the idea to directly compute the gradient in the {\em tangent space} to the unit sphere. In this work, in order to facilitate the numerical implementation, we first compute the gradient and then project it into the tangent space. For this purpose, it is very helpful to derive an explicit expression of the projected gradient that allows to preserve the unitary norm of the solution through the minimization procedure. It should be noted that explicit expressions of the projected gradient  are given in \cite{alouges} for the $\R^n$ gradient flow of the linear eigenvalue problem on the unit sphere and for the $L^2$ gradient flow of the Hartree-Fock nonlinear eigenvalue problem. We derive below an explicit expression of the projected gradient that allows to minimize trajectories when other Hilbert spaces than $L^2$ are considered.


Let $X = L^2$, $H^1,$ or $H_A$.  As previously stated, for each $u \in X$, one can find a member of $X$, denoted by $\nabla_X E(u)$, so that $E'(u)h = \langle h , \nabla_X E(u) \rangle_X$. We called such an element of $X$ a gradient of $E$ at $u$.  Consider $\beta : X \rightarrow \R$,
\begin{equation} \label{beta}
\beta(u) = \int_{\cal D} |u|^2.
\end{equation}
Since we want to minimize the energy $E(u)$ subject to the constraint $\beta(u)=1$, we obtain the {\em tangent space} for our problem:
\begin{equation}
  T_{u,X} = null ( \beta'(u)) = \{ w \in X: \langle u , w \rangle_{L^2} = 0 \}.
\end{equation}
Note that $T_{u,X}$ is a closed linear subspace of $X$, and for each $u \in X$, there exists a unique orthogonal projection from $X$ onto $T_{u,X}$.  We denote this projection by $P_{u,X}$.
Note also that $P_{u,X}$ is a linear transformation with domain $X$ and range $T_{u,X}$.  Thus, this transformation depends on the Hilbert space and $u \in X$.

Let $u_0 \in X$ so that $\beta(u_0)=1$ and write the descent method with the projected gradient:
\begin{equation} \label{sd}
z(0)=u_0 \text{ and } z'(t)=-P_{z(t),X} \nabla_X E(z(t)).
\end{equation}
We can easily see  that $\beta(z)$ is constant since
\begin{equation}
(\beta (z))'(t) = \beta'(z(t)) z'(t) = - \beta'(z(t)) (P_{z(t),X} \nabla_X E(z(t)))= 0,
\end{equation}
for all $t$, as $P_{z(t),X}$ is the projection of $X$ onto the null space of $\beta'(z(t))$.  Thus $\beta (z)$ is constant and  if $u=\lim_{t \rightarrow \infty} z(t)$, then $\beta(u) = \beta(u_0)$, and the norm of the initial state is preserved.  In conclusion, by projecting the Sobolev gradient of $E$ at $z(t)$ into the null space of $\beta'(z(t))$ for each $t$, we get that $z(t)$ satisfies the mass constraint for all $t$ (see \cite{jwn} for a more detailed development on this topic).

For numerical implementation purposes, we give below an heuristic derivation of the explicit expression of the projection (see \cite{pkme} for a more rigorous demonstration). If, for the sake of simplicity, ${\cal G} = \nabla_X E(u)$ denotes the Sobolev gradient gradient of $E$ at $u$, the projected gradient
is determined from the following two conditions:
\begin{equation}
P_{u,X} {\cal G} \in T_{u,X},
\label{eq-projg-1}
\end{equation}
\begin{equation}
\langle P_{u,X} {\cal G} , h \rangle_{X}  = E'(u)h, \, \forall h \in T_{u,X}.
\label{eq-projg-2}
\end{equation}
In order to satisfy \eqref{eq-projg-2}, we choose the projected gradient
of the form $P_{u,X} {\cal G} ={\cal G} - B v_X$, with $B\in \R$ a constant and $v_X \in X$ such as
\begin{equation}
\langle v_X , h \rangle_{X} = \langle u , h \rangle_{L^2}, \, \forall h \in X.
  \label{eq-vX}
\end{equation}
The constant $B$ is then obtained by imposing \eqref{eq-projg-1}. The final expression  that will be used for numerical implementation is:
\begin{equation}
  P_{u,X} {\cal G} = {\cal G} - \frac{\Re \langle u, {\cal G} \rangle_{L^2}}{\Re \langle u, v_X \rangle_{L^2}}\, v_X,
  \label{eq-projG}
\end{equation}
with $v_X$ computed from \eqref{eq-vX}. Note that if $X=L^2$, $v_X=u$ and we recover the explicit expression of the projected gradient given in \cite{alouges1}.
It is also important to note that, in regard to numerical consideration as well as obtaining global existence, uniqueness, and  asymptotic convergence, we need that the map $u \rightarrow P_{u,X} \nabla_X E(u)$ be $C^1$ as a map from $X$ to $X$.  Using the above expression for the projection, we  present in the next section some convergence results.

\subsection{Convergence results in an infinite dimensional Hilbert space}\label{subsec-theory}

In this subsection we define the evolution equation we use in the Hilbert space setting and give our global existence and convergence result for the constrained minimization problem. Here we extend the results obtained in \cite{pkme} for the case of the Gross-Pitaevskii energy without rotation.  In this work, as well as in \cite{pkme}, we move away from the general theory of Sobolev gradients as presented in \cite{jwn}, since the criteria for asymptotic convergence of the evolution equation for constrained minimization problems is not available in \cite{jwn}.

The idea below the following analysis is to show that the GP energy functional with rotation has the same properties as the GP energy without rotation if the norm $\| \cdot \|_{H_A}$ is used. We thus can adapt  the results obtained in \cite{pkme} to our case. We start by noting  that, due to the mass conservation, one can add a multiple of $\int_{\cal D} |u|^2$ to the Gross-Pitaevskii energy and the resulting functional will have the same minimizers as the original functional. The idea is to obtain a functional that is uniformly and strictly convex. We remind the reader that for $X$ a Hilbert space, we say that $E: X \rightarrow \R$ is uniformly and strictly convex if there exists $\epsilon > 0$ so that $E''(u)(h,h) \geq \epsilon |h|_X^2$ for all $h \in X$.

Indeed, let us consider the form \eqref{gpcov} of the energy functional, and  suppose that there exists $1 > \delta > 0$ so that $V_{eff} > \delta$. We observe that
\begin{eqnarray}
E''(u)(h,h) = \int_{\cal D} | \nabla_A h |^2 + 2V_{eff}|h|^2 + 2g(|u|^2 |h|^2 + 2(\Re \langle u, h \rangle))^2) \geq \\
\int_{\cal D} | \nabla_A h |^2 + 2V_{eff}|h|^2 \geq  \delta \int_{\cal D} | \nabla_A h |^2 + |h|^2 = \delta \|h\|^2_{H_A},
\end{eqnarray}
and infer that $E: H_A \rightarrow \R$ is uniformly and strictly convex with the assumption that $V_{eff}$ is bounded away from zero. Due to the equivalence of norms, $E: H^1 \rightarrow \R$ is also uniformly and strictly convex.  Note that if $V_{eff}$ is not bounded away from zero, then one can obtain this property by adding a multiple of the constraint to the energy.  This does not change the minimization problem as indicated by the following theorem.

\begin{theorem} \label{shift}
Let $E$ be a $C^2$ function on a subspace $X$ contained in $L^2({\cal D})$. Let
\begin{equation}
E_{\epsilon}(u) = E(u) + \epsilon \int_{\cal D} |u|^2.
\end{equation}
Then for $\beta(u)= \int_{\cal D} |u|^2$ and $h \in null(\beta'(u))$, $E'(u)h=0$ iff $E_{\epsilon}'(u)h=0$.
\end{theorem}

Some other properties of the functional are required to obtain the asymptotic convergence of the evolution equation.  In particular, we need the functional to be continuously twice Fr{\'e}chet differentiable and bounded from below.  The latter two properties are standard and we therefore omit them.  With these properties checked,  we can give our global existence and convergence in the asymptotic limit. Using the space $H_A$, the proof of the following two theorems are identical to the ones given in \cite{pkme}. Thus we omit the proofs and refer the reader to this work.

\begin{theorem} \label{thmsdexists}
Suppose $X$ is a Hilbert space and that  $E: X \rightarrow \R$ is continuously twice Fr{\'e}chet differentiable. Suppose also that $\beta:X \rightarrow \mathbb{R}$ is a given function such that, if $P_{u,X}$ denotes the orthogonal projection of $X$ onto the nullspace of $\beta'(u)$, then the map $u \rightarrow P_{u,X}$ is $C^1$. Then $z(t)$ given by \eqref{sd} is uniquely defined for all $t \geq 0$.
\end{theorem}
\begin{theorem} \label{asconv}
Suppose the hypothesis of Theorem \ref{thmsdexists} and that $z(t)$ is given by equation \eqref{sd}, with $\nabla_X E(u_0) \neq 0$.   If  $E: X \rightarrow \R$ is uniformly and strictly convex, then
\begin{equation}
\lim_{ t \rightarrow \infty} z(t) = u
\end{equation}
exists. Furthermore, there exist two constants $m$ and $c$ so that $\|u - z(t)\|_X \leq m e^{-ct}$, and $E'(u)h=0$ for all $h \in null (\beta'(u))$.
\end{theorem}

From the above two theorems, if we have in mind that
the functional $E$ defined in \eqref{gp} is continuously twice Fr{\'e}chet differentiable and uniformly and strictly convex when the domain is considered to be $H_A$ or $H$,  we obtain the result that $E$ has a minimizer in $H_A$ and in $H^1$ that satisfies the constraint $\beta$.  Furthermore, this minimizer is obtained as the limit of the trajectory we defined in \eqref{sd}.  This convergence result is not only important on its own, but, as we shall see, plays an important role in the rate of convergence  of our numerical  simulations.

\section{Numerical implementation} \label{sec-implement}

In this section we explain in detail the setup for our simulations using the descent method with both finite differences and finite elements discretization in two space dimensions. Both implementations follow the
general lines of the algorithm described below.

If $Y=L_G, H_G, H_{A_G}$ denotes the  finite dimensional Hilbert spaces resulting after the discretization of the domain ${\cal D}$,
the descent method \eqref{sd} takes the following discrete form:\\
starting from $u_0 \in Y$, define a trajectory $z_n, n \geq 1$ as (forward Euler scheme):
\begin{equation}
z_0=u_0, \quad z_{n+1} = z_n - \delta t_n \nabla_{z_n,Y} E_G(z_n),
\label{eq-algo-descent}
\end{equation}
where $\nabla_{u,Y} E_G (u)$ denotes the gradient obtained with respect to each inner product and projected  following \eqref{eq-projG}. The {\em time-step} value $\delta t_n$ could be optimized when computed as the local minimum of the real valued function
\begin{equation}
r \rightarrow E_G(z_n - r \nabla_{z_n,Y} E_G(z_n)).
\label{eq-algo-lines}
\end{equation}
As convergence criterion,  the algorithm stops when the relative change in energy $E_G$ is below of an imposed limit. {
We note that in the continuous steepest descent algorithm, the constraint was satisfied for all time $t$ and hence for the  converged solution.  In the discrete case, due to the first order discretization in time, it is easy to see from \eqref{eq-algo-descent} that the norm is conserved at time level $(n+1)$ up to an error of order $(\delta t_n)^2 \| \nabla_{z_n,Y} E_G(z_n) \|_{L^2}$.
After each (or several) iteration(s), one could also normalize the solution, as in \cite{pierre} where a Sobolev descent method with step-size 1 is used.  This results in an improvement in the accuracy to which the constraint is preserved.  The main observation that we made was that even though we used a first order discretization in time, our projection method allowed to take larger time steps when compared to the method using the normalization alone.}

\newpage

\subsection{Finite differences}

We discretize $\cal D$ into an $N$ by $N$ equally spaced ($\delta_x = \delta_y =\delta$) grid and let ${\cal D}_G$ be the set of all $K=N^2$ grid points. Let $X$ be the collection of all complex valued functions on $D_G$. For $f \in X$, $(D_1 f)(x,y)$ is the approximation to the partial derivative in the first independent variable at $(x,y)$ and $(D_2 f)(x,y)$ is the approximation to the partial derivative in the second independent variable at $(x,y)$.
We have used a fourth-order centered finite-difference scheme to approximate the first partial derivatives. When compared to the classical second order scheme, this high-order approximation proved very helpful in computing complex configurations (vortex lattices within the condensate) with reasonably fine grids.
Furthermore $Df = \binom{D_1f}{D_2f}$.  For $(x,y)$ a grid point, we also define $D_{1,A}$ and $D_{2,A}$ by
\begin{equation}
(D_{1,A} f)(x,y)= (D_1f) (x,y) + i \Omega y f(x,y)
\end{equation}
and
\begin{equation}
(D_{2,A} f)(x,y)= (D_2f) (x,y) - i \Omega x f(x,y).
\end{equation}
We denote by $D_A = \binom{D_{1,A}}{D_{2,A}}$, the discretized form of the operator  $\nabla_A $.  The three inner products that equip $X$ are defined as: for $f,g \in X$,
\begin{equation}
\langle f , g \rangle_{L^2} = \langle f , g \rangle,
\end{equation}
\begin{equation}
\langle f , g \rangle_{H} = \langle f ,g \rangle + \langle D_1f , D_1g \rangle + \langle D_2f , D_2g \rangle,
\end{equation}
and
\begin{equation}
\langle f ,g \rangle_{H_A} = \langle f , g \rangle + \langle D_{1,A}f , D_{1,A}g \rangle+ \langle D_{2,A}f , D_{2,A}g \rangle,
\end{equation}
where $\langle \cdot , \cdot \rangle$ denotes the complex $\C^K$ inner product.  Note that $\langle \cdot , \cdot \rangle_{L^2}$ is analogous to the $L^2({\cal D}, \C)$ inner product,  $\langle \cdot, \cdot \rangle_H$ is analogous to the $H^{1,2}({\cal D}, \C)$ inner product, and $\langle \cdot , \cdot \rangle_{H_A}$ is analogous to the $H_A({\cal D},\C)$ inner product.

Since $D_1, \ D_2, \ D_{1,A}, \ D_{2,A}$ can be viewed as a linear transformation acting on $\mathbb{C}^K$, we think of each of these transformations as a $K \times K$ matrix.  Let $D_M^*$ denote the conjugate transpose of the corresponding matrix. We note that we can write the $H$ and $H_A$ inner products as
\begin{equation} \label{gradH}
\langle f , g \rangle_H = \langle (I + D_1^*D_1 + D_2^*D_2) f , g \rangle_{L^2}.
\end{equation}
and
\begin{equation} \label{gradHA}
\langle f , g \rangle_{H_A} = \langle (I + D_{1,A}^*D_{1,A} + D_{2,A}^*D_{2,A}) f , g \rangle_{L^2}.
\end{equation}
The collection $X$ makes a finite dimensional Hilbert space with each of the above inner products.  We denote the resulting Hilbert spaces by $L^2_G$, $H_G$, and $H_{A_G}$.  Now, we discretize the energy functional as given in equations \eqref{gp} and \eqref{gpcov}.  Here the subscript $G$ denotes that we are in the finite difference setting.
\begin{eqnarray}
E_G(f)= \delta^2  \sum_{{\cal D}_G} \frac{1}{2} (|D_1 f|^2 + |D_2 f|^2) + V_{trap_G}|f|^2 + \frac{g}{2}|f|^4 - \Omega\, rot_G f,
\end{eqnarray}
where for $\x \in {\cal D}_G$,
\begin{equation}
rot_G f(\x) =  \Re \ (i f(\x)^*A(\x) \binom{(D_1 f)(\x)}{(D_2 f))(\x)},
\end{equation}
and
\begin{equation}
A(x,y) = (y \ \ -x).
\end{equation}
Equivalently,
\begin{eqnarray}
E_G(f)= \delta^2 \sum_{ {\cal D}_G}  \frac{1}{2} \left( |D_{1,A}f|^2 + |D_{2,A}|^2 \right) + V_{eff_G}|f|^2 + \frac{g}{2}|f|^4.
\end{eqnarray}
If we take a derivative of $E_G$, we get that
\begin{eqnarray}\nonumber
E_G'(f)h= \delta^2 \Re \sum_{ {\cal D}_G}  \ \langle D_1h , D_1f \rangle &+& \langle D_2h , D_2f \rangle \\
&+&
 2\langle h , V_{trap_G} f + g |f |^2 f - \Omega i A^t (Df)\rangle.
\end{eqnarray}
Observe that for each $f \in X$, $E_G'(f)$ is a continuous linear transformation on $X$ using any of the three norms we specified.  Thus it has a representation with respect to each of the inner products we defined above.  Using this representation, we will obtain a gradient.
Since the $L^2$ inner product is proportional to the Euclidean inner product, the ordinary or Euclidean gradient (i.e. the list of partial derivatives of $E_G$ taken with respect to each of the $K$ independent variables) is easily derived if the real valued transformation $E_G'(f)$ is rewritten as:
\begin{equation} \label{gradh}
E_G'(f)h=  \Re \langle h , \nabla_{L^2} E(f) \rangle _{L^2}.
\end{equation}
We get that
\begin{eqnarray}
\nabla_{L^2} E(f)= \delta^2( D_1^*D_1 f\ + \ D_2^*D_2 f\ + 2(V_{trap_G}f \ + g |f|^2 f\ -\Omega i A^t(Df))).
\end{eqnarray}

We now derive the other two gradients, $\nabla_H E(f)$ and $\nabla_{H_A} E (f)$,  with respect to the $H$ and $H_A$ inner products.  From \eqref{gradH} and \eqref{gradHA} we obtain that
\begin{equation}
E_G'(f)h = \Re \langle h , \nabla_H E_G (f) \rangle_H = \Re \langle h , ( I+D^*D)\nabla_H E_G (f) \rangle_{L^2},
\end{equation}
and
\begin{equation}
E_G'(f)h = \Re \langle h , \nabla_{H_A} E_G (f) \rangle_{H_A} = \Re \langle h , (I+D_A^*D_A) \nabla_{H_A} E_G (f) \rangle_L.
\end{equation}
By comparing these equations to \eqref{gradh}, we finally get that
\begin{equation} \label{gH}
\nabla_H E_G (f) = (I+D^*D)^{-1} \nabla_{L^2} E(f),
\end{equation}
and
\begin{equation} \label{gHA}
\nabla_{H_A} E_G (f) = ( I+D_A^*D_A)^{-1} \nabla_{L^2} E(f).
\end{equation}

The discrete descent method \eqref{eq-algo-descent} using the above  finite difference dscretization was implemented in Matlab.  The Sobolev gradients are computed  from \eqref{gH} and \eqref{gHA} by solving linear systems at each time step using a preconditioned conjugate gradient method. Since this part is time consuming on fine grids, we used a linesearch algorithm to locally compute the time step from \eqref{eq-algo-lines}. This resulted in a significant reduction of the number of iterations needed to achieve convergence.

\subsection{Finite elements}

The finite-elements implementation uses the free software FreeFem++ \cite{freefem}, which proposes a large variety of triangular finite elements (linear and quadratic Lagrangian elements, discontinuous $P1$, Raviart-Thomas elements, etc.) to solve partial differential equations (PDE) in two dimensions (2D). FreeFem++ is an integrated product with its own high level programming language with a syntax close to mathematical formulations.

It is therefore very easy to implement the variational formulations associated to the calculation of the three gradients, since the definitions of scalar products \eqref{eq-innpL}--\eqref{eq-innpHA} use an integral form. Following the developments in section \eqref{subsec-grad}, and using as definition of the complex inner product $\langle u, v \rangle = u v^*$, the ordinary gradient is derived from \eqref{derivgp} and computed as the solution ${\cal G}=\nabla_{L^2} E$ of the problem with homogeneous Dirichlet boundary conditions:
\begin{eqnarray}
\int_{\cal D}  {\cal G} \, h  &=& \mathrm{RHS}, \label{eq-FE-gl2}\\
\mathrm{RHS} &=& \int_{\cal D}  \nabla u  \nabla h  + 2\left[ V_{trap}\, u + (g |u|^2 )u  - i \Omega  A^t\nabla u \right] h,
\label{eq-FE-rhs}
\end{eqnarray}
where $h$ stands now for the real valued basis function of the finite element space.
Following \eqref{el}, the $H^1$ gradient is directly computed by solving the equation:
\begin{equation}\label{eq-FE-GH1}
\int_{\cal D}  \nabla {\cal G} \, \nabla h  + {\cal G}  h  = \mathrm{RHS}, \quad\mbox{where}\quad  {\cal G} = \nabla_{H} E.
\end{equation}
It is interesting to note that \eqref{eq-FE-GH1} is directly derived from  the weak formulation of \eqref{el}, with the obvious advantage to
obtain a simpler right-hand side \eqref{eq-FE-rhs}, which is derived by integrating by parts the weak form of the  $L^2$ gradient. Therefore, in order to solve \eqref{eq-FE-GH1}, it is not necessary to explicitly compute the $L^2$ gradient (by solving \eqref{eq-FE-gl2}), as required for the finite-difference implementation.

Observing from \eqref{HAidentity} that the $H_A$ scalar product could be expanded to obtain the equivalent definition:
\begin{equation} \label{eq-FE-HA}
<u, v> _{H_A} = \int_{\cal D} \langle \left[ 1 + \Omega^2 (y^2+x^2) \right]  u , v \rangle + \langle \nabla u , \nabla v \rangle -2i\Omega \langle A^t\nabla u , v \rangle,
\end{equation}
the $H_A$ gradient is directly computed as the solution ${\cal G} = \nabla_{H_A} E$ of the problem:
\begin{equation}\label{eq-FE-GHA}
\int_{\cal D}  \left[ 1 + \Omega^2 (y^2+x^2) \right]  {\cal G} h +  \nabla {\cal G}  \nabla h -2i\Omega  (A^t\nabla {\cal G}) h =  \mathrm{RHS}.
\end{equation}
It is interesting to emphasize the fact that previous equations are solved in complex variables. The approach based on the separation of the real and imaginary part of the gradient used in \cite{lookman} is not possible when computing the $H_A$ gradient. The FreeFem scripts are written in an optimized form using the pre-computation and factorization of the complex matrices associated to linear systems given by (\ref{eq-FE-GH1}) and (\ref{eq-FE-GHA}).
It is interesting to note that the same matrices are involved in the computation of $v_X$ from \eqref{eq-vX}; the projected gradient \eqref{eq-projG} could be therefore optimized in the same way.
The implementation uses P1 (piecewise linear) finite-elements, with a P4 representation of the nonlinear terms appearing in \eqref{eq-FE-rhs}. A fifth order quadrature formula was used to compute two-dimensional integrals. The FreeFem scripts allow
to switch to P2 (piecewise quadratic) finite elements by a simple change of the definition of the generic finite-elements space. Adaptive mesh refinement was used for simulations of rotating BEC with dense lattice of vortices.

\section{Numerical experiments} \label{numerics}

We first use a test case with analytical manufactured solution to ascertain the convergence of the steepest descent method for each of the three gradients. Then, we use the numerical set-up to compute simple metastable states of rotating Bose-Einstein condensates with single or multiple vortices. The performances of the three methods are comparatively evaluated. Finally, the new $H_A$ gradient method is used to compute complex configurations relevant for real rotating condensates (Abrikosov vortex lattice and giant vortex).

\subsection{Test case with manufactured solutions}

This test case is used as benchmark for the evaluation of the descent method for each of the three gradients ($L^2, H, H_A$).
We consider a non-linear problem close to the Gross-Pitaevskii equation:
\begin{equation}\label{eq-FF-testC-eq}
 -\frac{1}{2} \nabla^2 u + C_{trap}\, u + g |u|^2 u -i\Omega (A^t\nabla)u = f,
\end{equation}
corresponding to the minimization of the energy functional:
\begin{equation} \label{eq-FF-testR-en}
E(u,f) =
\int_{\cal D}  \frac{1}{2}|\nabla u|^2
+ C_{trap} \,|u|^2+ {g}\,\frac{|u|^4}{2} -(f^* u + f u^*)
 - \Omega \Re(iu^* A^t \nabla u).
\end{equation}
For this energy functional, the $L^2$ gradient is expressed as in \eqref{el}, with $V_{trap}=C_{trap}=const.$ and a supplementary term $-2\langle f, h\rangle$ to be added. It should be noted that this is a test case of minimization without constraint.

\begin{figure}[!h]
\centering
\includegraphics[width=0.7\columnwidth]{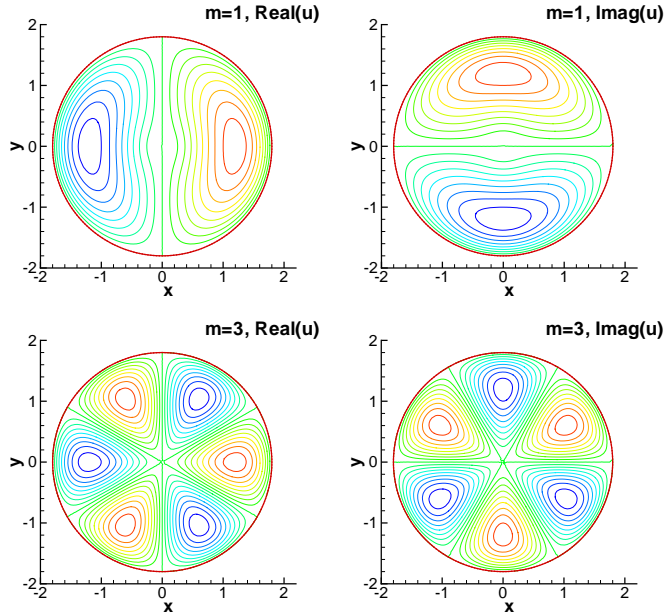}
\caption{Contour patterns of the manufactured solution corresponding to equations \eqref{eq-FF-testC-sol} and \eqref{eq-FF-testC-solU}. Azimuthal wave numbers $m=1$ and $m=3$.}
\label{fig-manuf}
\end{figure}

In order to test the implemented methods, we {\em manufacture} solutions of (\ref{eq-FF-testC-eq}): we consider a given expression for $u$ and calculate the corresponding right-hand side $f(x,y)$.
A simple way to construct such manufactured solutions is to consider solutions with azimuthal symmetry:
\begin{equation}\label{eq-FF-testC-sol}
u_f(x,y) = U(r) \, \exp(i m \theta),
\end{equation}
where $(r,\theta)$ are cylindrical coordinates ($r=\sqrt{x^2+y^2}$). Since the Laplacian in cylindrical coordinates reads
\begin{equation}
\nabla^2 = \frac{1}{r} \frac{\pl}{\pl r}\left(r \frac{\pl}{\pl r}\right) +
\frac{1}{r^2} \frac{\pl^2}{\pl \theta^2},
\end{equation}
and the new term corresponding to the rotation becomes
\begin{equation}
A^t \nabla u = y\frac{\pl u}{\pl x} - x \frac{\pl u}{\pl y} = - \frac{\pl u}{\pl \theta}.
\end{equation}
we obtain that
\begin{equation}
f = F(r) \, \exp(i m \theta),
\end{equation}
with
\begin{equation}\label{eq-FF-testC-fr}
F(r) = -\frac{1}{2} \frac{1}{r} \frac{\pl}{\pl r}\left(r \frac{\pl U}{\pl r}\right)
+ \frac{1}{2} \frac{m^2}{r^2} U +  C_{trap}\, U + g U^3 - m \Omega U.
\end{equation}
We choose the domain ${\cal D}$ to be a circle of radius $R$ and
\begin{equation}
U = r^2(R-r),
\label{eq-FF-testC-solU}
\end{equation}
which satisfies the homogeneous boundary condition $u=0$ for $r=R$.
For this choice, we obtain useful analytical formulas for
\begin{equation}
F(r)= -\frac{1}{2}(4R-9r) + \frac{1}{2} m^2(R-r) + C_{trap}\, U + C_N U^3 - m \Omega U,
\end{equation}
and energy
\begin{equation} \label{eq-FF-testC-enaU}
E(u,f) = 2\pi\left(-\frac{R^6}{20}-m^2 \frac{R^6}{120} -C_{trap} \frac{R^8}{168} -C_N  \frac{3R^{14}}{20020} \right) +  m \Omega \pi \frac{R^8}{84}.
\end{equation}
The contour patterns for such solutions are displayed in Fig. \ref{fig-manuf} for $m=1$ and $m=3$.

The numerical application for manufactured solutions consider the following parameters:
\[
C_{trap}=20, \quad g=100, \quad R=1, \quad m=3, \quad \Omega=10.
\]
For this case, the theoretical values for energy and angular momentum of the exact solution are:
$E = -0.505553$ and $L_z = 0.1122$, respectively. The computation  is considered as converged if the relative variation of the energy is less than $\varepsilon=10^{-8}$.

Tables \ref{tab-manuf-FD} and \ref{tab-manuf-FE} assess  the convergence of the descent method by computing different norms of the difference between the exact and computed solutions. Performance of each gradient method are quantified by extracting the overall computing (CPU) time and the number $n$ of time steps necessary to achieve convergence.

All test cases considered $u_0=0$ as the initial guess for the descent method. Different initial conditions (e.g. $u_0$ computed as the solution of the corresponding linear problem) were tested with similar convergence results.

\begin{table}[!h]
\centering
\begin{tabular}{ | l | l | l | l | l | l | l | l | l |} \hline
    &   $N$ &       $n$     &   CPU    & E(u) &   $\|u-u_f\|_{\infty}$ & $\|u-u_f\|_{L^2}$ & $\|u-u_f\|_{H}$ \\
\hline \hline
    $L$~ ~  &   $2^6$~ ~  & 1137~~  &   35.41~~ &   -.4828~~&   .0010~~&    2.25e-4~~&  .0270   \\
\hline
    $H$ &   $2^6$   &   100 &   16.23   &   -.4828&     2.81-4  &   1.66e-5&    .0021   \\
\hline
    $H_A$   &   $2^6$   &   38  &   6.39    &   -.4828&     1.41e-4 &   3.85e-6&    4.73e-4\\
\hline \hline
    $L$ &   $2^7$   &   1960    &   308.13  &   -.4941&     4.11e-4 &   2.84e-4&    .0185\\
\hline
    $H$ &   $2^7$   &   40  &   36.34   &   -.4941&     6.32e-5 &   4.81e-4&    5.28e-4\\
\hline
    $H_A$   &   $2^7$   &   18  &   17.22   &   -.4941&     2.05e-5 &   6.84e-7&     6.26e-5\\
\hline \hline
    $L$ &   $2^8$   &   $>3000$ &   $>2.04e3$&  -.4985&         &       &       \\
\hline
    $H$ &   $2^8$   &   30  &   154.47  &   -.4997&     4.59-5  &   9.77e-6&    .0013 \\
\hline
    $H_A$   &   $2^8$   &   14  &   73.21   &   -.4997&     1.56e-6 &   1.36e-6&     1.504e-4\\
\hline
\end{tabular}
\caption{Test case with manufactured solutions. Algorithm efficiency and convergence test for the finite difference implementation (variable time step computation).}
\label{tab-manuf-FD}
\end{table}

\begin{table}[!h]
\centering
\begin{tabular}{ | l | l | l | l | l | l | l | l | l |l|} \hline
 &      $M$/Triangles   &       $n$     &   CPU    & E(u) &   $\|u-u_f\|_{\infty}$ & $\|u-u_f\|_{L^2}$ & $\|u-u_f\|_{H}$ & $\delta t$\\
\hline \hline
    $L$~ ~  &   100/1776~ ~  &  1176~~  &   85~~    &   -.4934~~&   1.988e-3~~&     1.001e-3~~& 1.705e-2 & 8e-4 \\
\hline
    $H$ &   100/1776&   47  &   3.4 &   -.4934&     1.883e-3&   9.220e-4&  1.668e-3    & 1\\
\hline
    $H_A$   &   100/1776&   14  &   1   &   -.4934&     1.880e-3&   9.140e-4&   1.665e-2 & 3\\
\hline \hline
    $L$ &   200/7064&   4292    &   1252    &   -.5025&     7.492e-4&   4.200e-4&   7.401e-3 & 2e-4\\
\hline
    $H$ &   200/7064&   47  &   13.8    &   -.5025&     5.530e-4&   2.232e-4&  6.548e-3 & 1\\
\hline
    $H_A$   &   200/7064&   14  &   4.1 &   -.5025&     5.390e-4&   2.119e-4&   6.474e-3 & 3\\
\hline \hline
    $L$ &   400/27604&  $>8000$ &   $>9193$&    -.5027&     &   &    & 5e-5 \\
\hline
    $H$ &   400/27604&  47  &   54.2    &   -.5047&     1.687e-4&   6.8535e-5&  3.954e-3 & 1\\
\hline
    $H_A$   &   400/27604&  14  &   16.2    &   -.5047&     1.549e-4&   5.730e-5&   3.791e-3 & 3\\
\hline
\end{tabular}
\caption{Test case with manufactured solutions. Algorithm efficiency and convergence test for the finite element implementation (fixed time step computation). The triangular mesh is generated with $M$ points on the border of the domain.}
\label{tab-manuf-FE}
\end{table}

The first obvious observation is that the descent method using the ordinary $L^2$ gradient has very slow convergence rate because of  very small time steps imposed by the stability limit of the method. This was expected since this method is the equivalent to the explicit Euler integration scheme for the imaginary-time propagation equation. A similar result was reported in \cite{lookman} for simpler Schr{\"o}dinger type energy functionals. Larger time steps are allowed in the $H^1$ and $H_A$ methods, since the Sobolev gradients represent a preconditioning of the ordinary gradient \cite{garcia-ripoll,pierre,alouges1}.

For the descent methods using a constant time step $\delta t$ (finite element implementation), we compare the computations performed using the maximum value $(\delta t)_{max}$ allowed by the stability of each method.
These values, displayed in Tab.  \ref{tab-manuf-FE}, were  obtained by successive tests: the value of $\delta t$ was increased by 20\% for each new run, until the computation became unstable. It should be noted that we were not interested in a refined numerical evaluation of the stability limit of each method, since computations using a more precise estimation of $(\delta t)_{max}$ did not result in a significant variation of the CPU time. The same approach to compare methods using their maximum time step allowed by stability reasons will be applied to all subsequent computations in this section.

Tables \ref{tab-manuf-FD} and \ref{tab-manuf-FE} also allows to relate the computing cost to the complexity of each method. As already stated, the descent method using the $L^2$ gradient can be regarded as an explicit backward Euler scheme. It therefore has little complexity and the computing cost per iteration step ({\em i.e.} the ratio CPU/n) is very low.  Sobolev gradients are computing by solving linear systems, which adds extra computational cost. For the finite-difference implementation, equations \eqref{gH} or \eqref{gHA} are solved by a preconditioned conjugate-gradient method; since this part of the algorithm is time consuming, the CPU time per iteration step  (CPU/n) is multiplied up to a factor of 8, when compared to the $L^2$ gradient method. The situation is different in the finite-element implementation. Since the weak formulation of the equation \eqref{el} is used, the computation of all gradients needs to solve a linear system. In order to have an optimized numerical implementation that can switch between the three descent methods, the matrix of this system is stored and factorized before the time loop. As a consequence,
even though the matrix of the system in \eqref{eq-FE-gl2} is simpler (mass matrix) than in \eqref{eq-FE-GH1} or \eqref{eq-FE-GHA}, the ratio (CPU/n) is identical for the computation off all gradients.

In all numerical tests, the convergence of the $L^2$ gradient method needs a large number of time steps, and, consequently, much larger CPU times than the Sobolev gradient methods. Since the performances of the $L^2$ gradient method are very poor, it will not be used in the following numerical experiments. We shall now focus on the comparison between the  $H$ and $H_A$ method.  For this test case considering a large value of
$\Omega$, the $H_A$ gradient allows for larger time steps and therefore the computational time is considerably reduced, by approximately a factor of 3.
This suggests that the preconditioning of the gradient introduced by the new $H_A$ inner product is very effective for computing cases with high rotation frequencies $\Omega$ (it goes without saying that the $H$ and $H_A$ methods are equivalent for $\Omega \rightarrow 0$).

\subsection{Simulations of rotating Bose-Einstein condensates}

In computing stationary states of rotating Bose-Einstein condensates, the initial state  $u_0$ in the descent method \eqref{eq-algo-descent} plays a crucial role. The algorithm usually starts from a wave function distribution derived from the Thomas-Fermi approximation. In the strong interaction regime (large values of $g$), it is reasonable to neglect the contribution of the kinetic energy and work with the simplified energy functional:
\begin{equation}
E_{\TF} (\rho) = \int_{\cal D} V_{trap}\rho + \frac{g}{2} |u|^4.
\end{equation}
The minimizer of this energy corresponds to the Thomas-Fermi atomic density:

\begin{equation}
\rho_{\TF}(r) = |u|^2 =\left(\frac{\mu - V_{trap}}{g}\right)_+,
\label{eq-rhoTF}
\end{equation}
where $\mu$ is the chemical potential. Since $\mu$ is a Lagrange multiplier, imposing the mass constraint in \eqref{eq-rhoTF} yields a relation for $\mu$.
After computing the value of $\mu$, the  Thomas-Fermi radius of the condensate can be determined from \eqref{eq-rhoTF} ($\rho_{\TF}(R_{TF})=0$). When a rotation $\Omega$ is applied, the Thomas-Fermi approximation \eqref{eq-rhoTF} stands with $V_{eff}$ replacing $V_{trap}$. The resulting radius $R_{\TF}^\Omega$ is
used to estimate the size of the domain $\cal D$ in simulations ($r_D > R_{\TF}^\Omega$) .

We also mention that the converged final state is characterized by its energy $E(u)$ and angular momentum $L_z(u)$ which gives a measure of the rotation:
\begin{equation}
   L_z(u) = \int_{\cal D} \Re \left(i u^* (A^t \nabla) u \right).
  \label{eq-LZ}
\end{equation}

\subsubsection{Off-center vortex case: harmonic trapping potential and small $\Omega$}

\begin{figure}[!h]
\centering
\includegraphics[width=0.40\columnwidth]{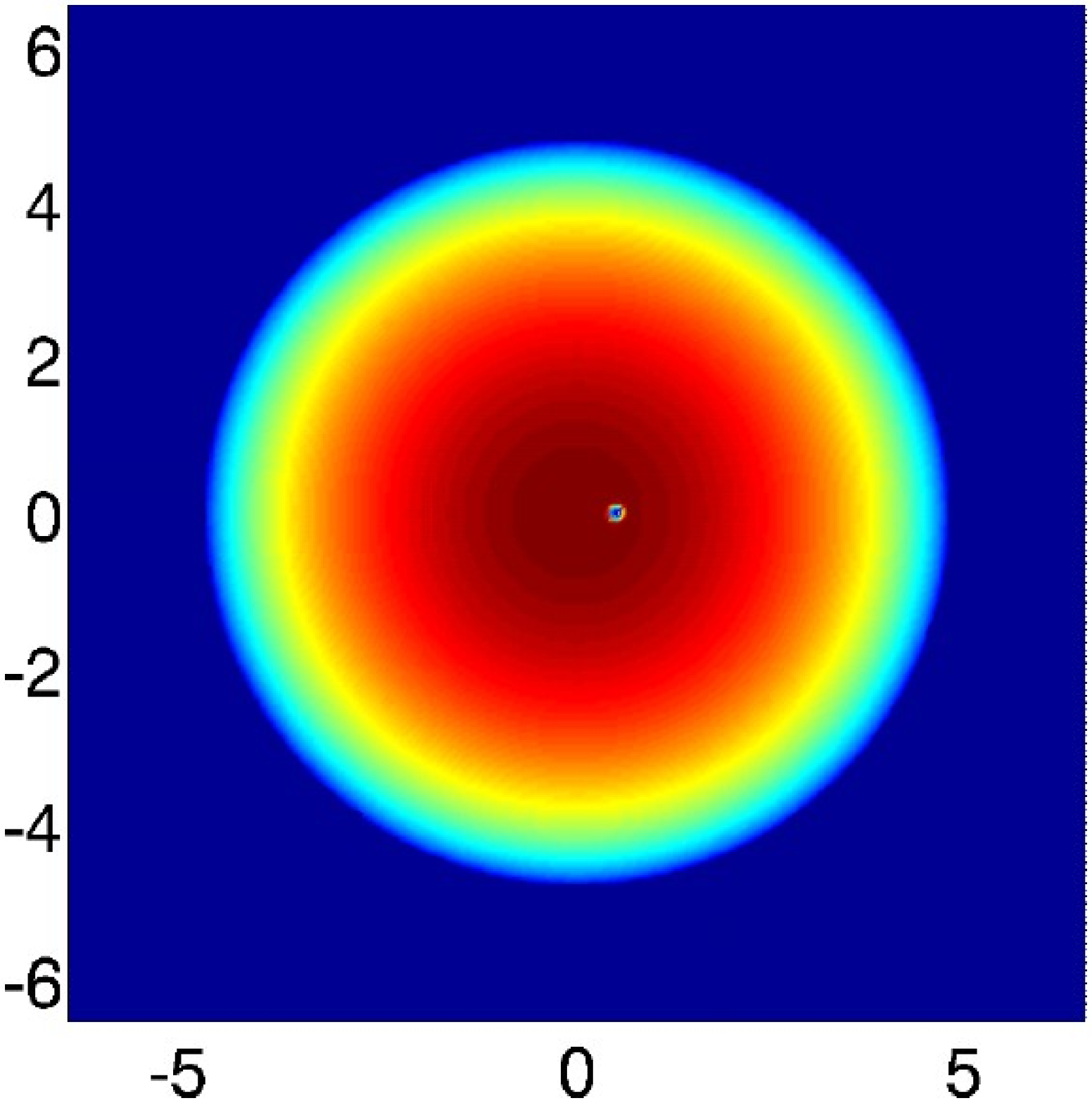}
\includegraphics[width=0.40\columnwidth]{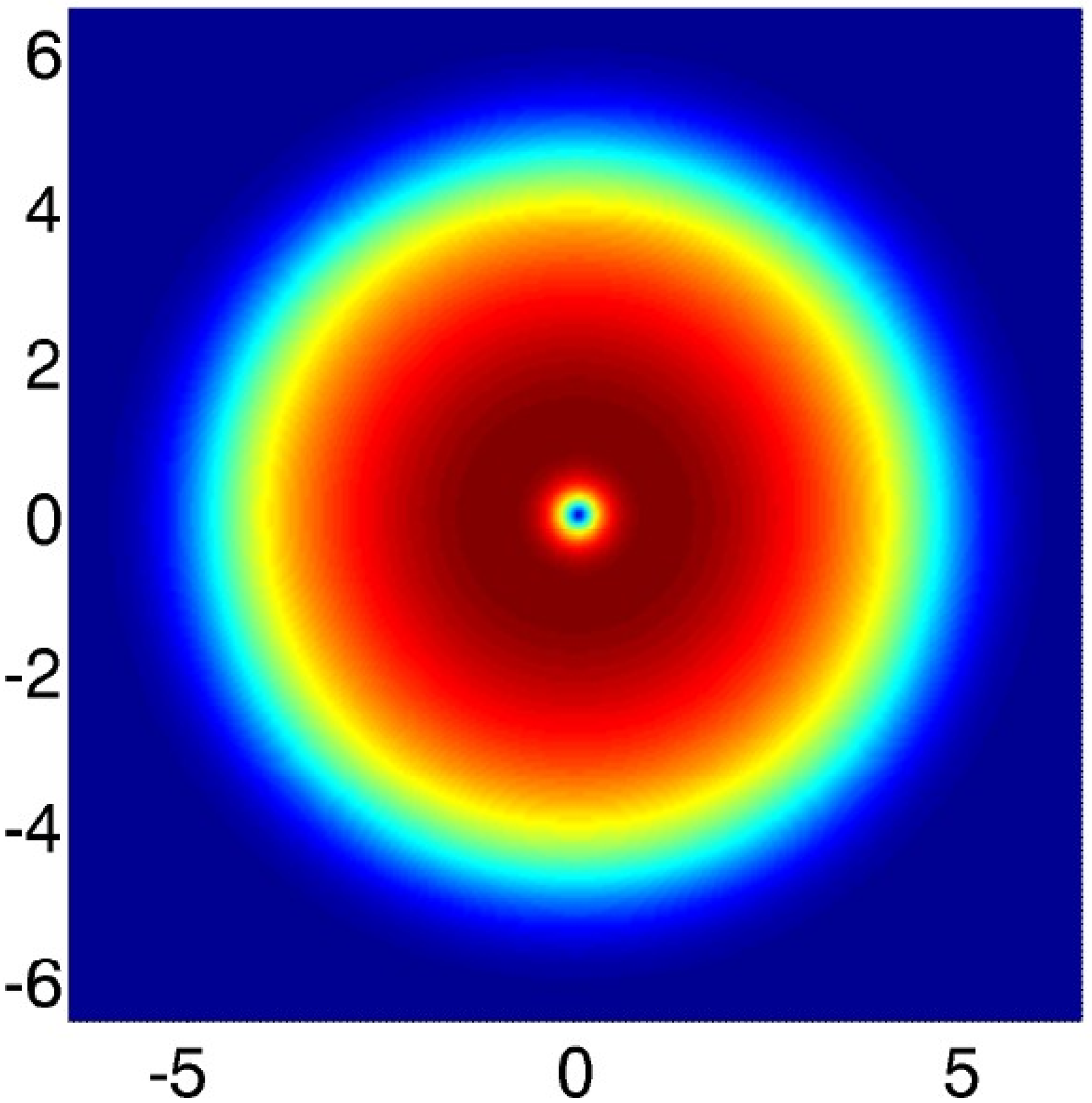}
\caption{Off-center vortex case. Initial state with an off-center vortex and final converged state with a centered vortex. Contours of atomic density $|u|^2$.}
\label{fig-offcenter}
\end{figure}

The second numerical experiment considers the classical harmonic trapping potential and an initial state computed from the Thomas-Fermi approximation plus a singly quantized vortex of center located at $(x_v, y_v)$. We use an ansatz for the vortex described in \cite{danaila1}. The parameters of the simulation are the following:
\begin{equation}
g = 500, \quad V_{trap}=r^2/2, \quad \Omega = 0.4, \quad x_{v}=0.5, \quad y_{v}=0.
\end{equation}
The Thomas-Fermi radius is for this case
$ R_\TF^{\Omega} = 5.246$
and the computational domain is circular of radius $R= 1.25 R_\TF^\Omega=6.56$. The final converged state contains a single vortex centered at the origin (see Fig. \ref{fig-offcenter}).

\begin{table}[!h]
\centering
\begin{tabular}{ | l | l | l | l | l | l | } \hline
 &      $N$ &       $n$     &   CPU    &    E(u)    &       $L_z(u)$   \\
\hline \hline
    $H$ &   $2^6$   &   1313    &   169.51  &   8.3587  &   .9998 \\
\hline
    $H_A$   &   $2^6$   &   1197    &   166.34  &   8.3587  &   .9998   \\
\hline \hline
    $H$ &   $2^7$   &   1184    &   866.88  &   8.3605  &   .9999   \\
\hline
    $H_A$   &   $2^7$   &   1127    &   890.06  &   8.3605  &   .9999   \\
\hline \hline
    $H$ &   $2^8$   &   1274    &   4.9548e3&   8.3606  &   .9999       \\
\hline
    $H_A$   &   $2^8$   &   1244    &   4.7882e3&   8.3606  &   .9999   \\
\hline

\end{tabular}\\
\caption{Off-center vortex case.  Algorithm efficiency and characterization ($E(u), L_z(u)$) of the converged state  state  for the finite difference implementation (variable time step computation).}
\label{tab-exp2-FD}
\end{table}

\begin{table}[!h]
\centering
\begin{tabular}{ | l | l | l | l | l | l | } \hline
&   $M$/triangles   &       $n$     &   CPU    &    E(u)    &       $L_z(u)$   \\
\hline \hline
    $H$ &   100/1762    &   701 &   56.97   &   8.3819  &   .994598 \\
\hline
    $H_A$   &   100/1762    &   703 &   57.47   &   8.3795  &   .994575 \\
\hline \hline
    $H$ &   200/7064    &   1667    &   537.85  &   8.3720  &   1.00042 \\
\hline
    $H_A$   &   200/7064    &   1717    &   556.41  &   8.3694  &   1.00022 \\
\hline \hline
    $H$ &   400/27604   &   1788    &   2.335e3&    8.3699  &   1.00052 \\
\hline
    $H_A$   &   400/27604   &   1831    &   2.407e3&    8.3673  &   1.00032 \\
\hline
\end{tabular}
\caption{Off-center vortex case.  Algorithm efficiency and characterization ($E(u), L_z(u)$) of the converged state  state for the  finite element implementation. The time step  is set to 0.1 for all computations.}
\label{tab-exp2-FE}
\end{table}

The comparative results are presented in Tabs. \ref{tab-exp2-FD} and \ref{tab-exp2-FE}.  It is important to note that the convergence test must be set to $\varepsilon=10^{-8}$ in order to obtain a final state with a vortex centered at the origin and $L_z=1$ (theoretical value reached for the finest meshes). A relaxed convergence criterion will result in a vortex that is not exactly centered since the convergence rate is very slow at the end of the simulation. As expected, the $H_1$ and $H_A$ perform similarly because of the low value of $\Omega$.

\pagebreak

\subsubsection{Vortex array case: harmonic-plus-quartic trapping potential and large $\Omega$}

\begin{figure}[!h]
\centering
\includegraphics[width=0.40\columnwidth]{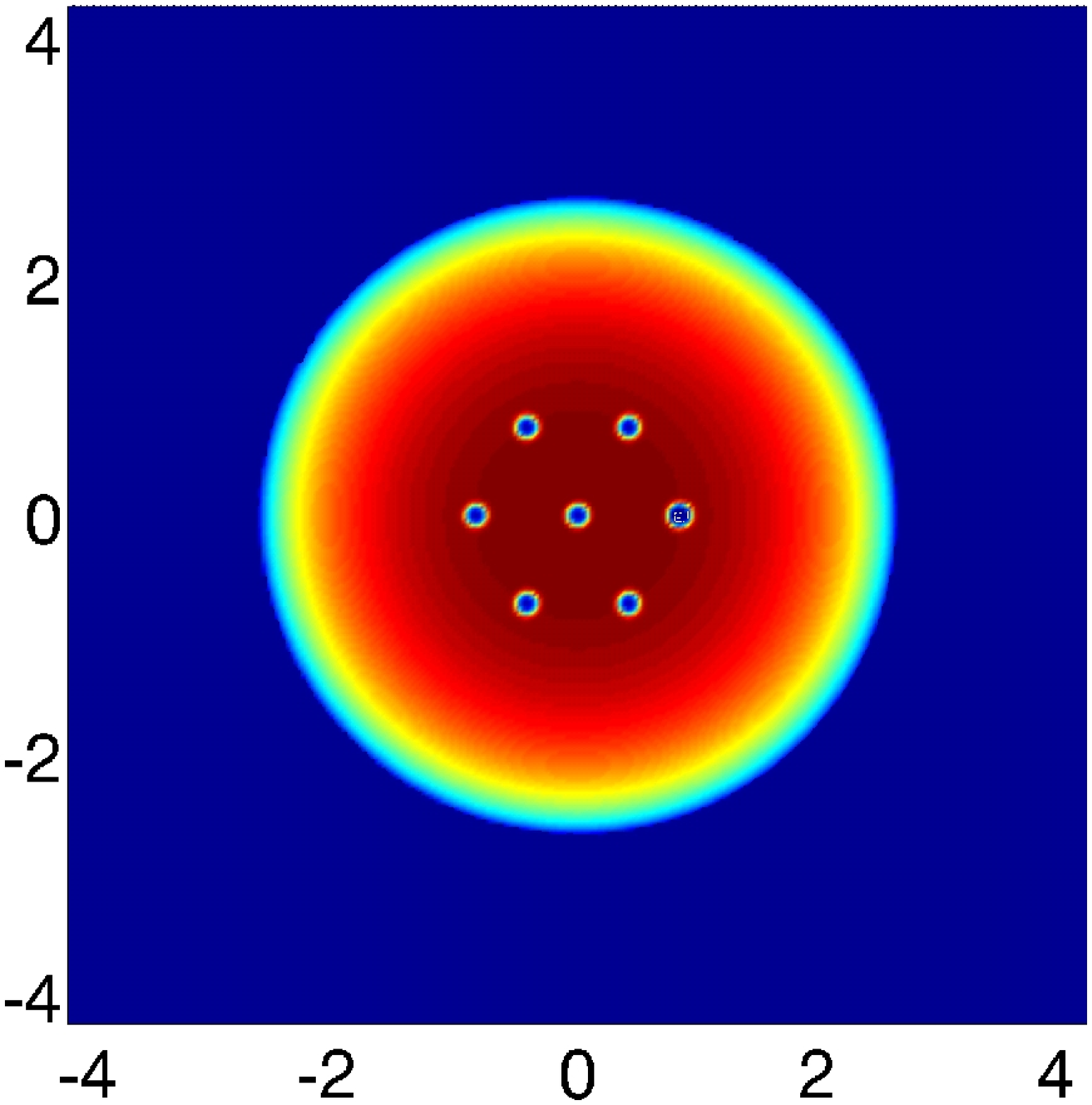}
\includegraphics[width=0.40\columnwidth]{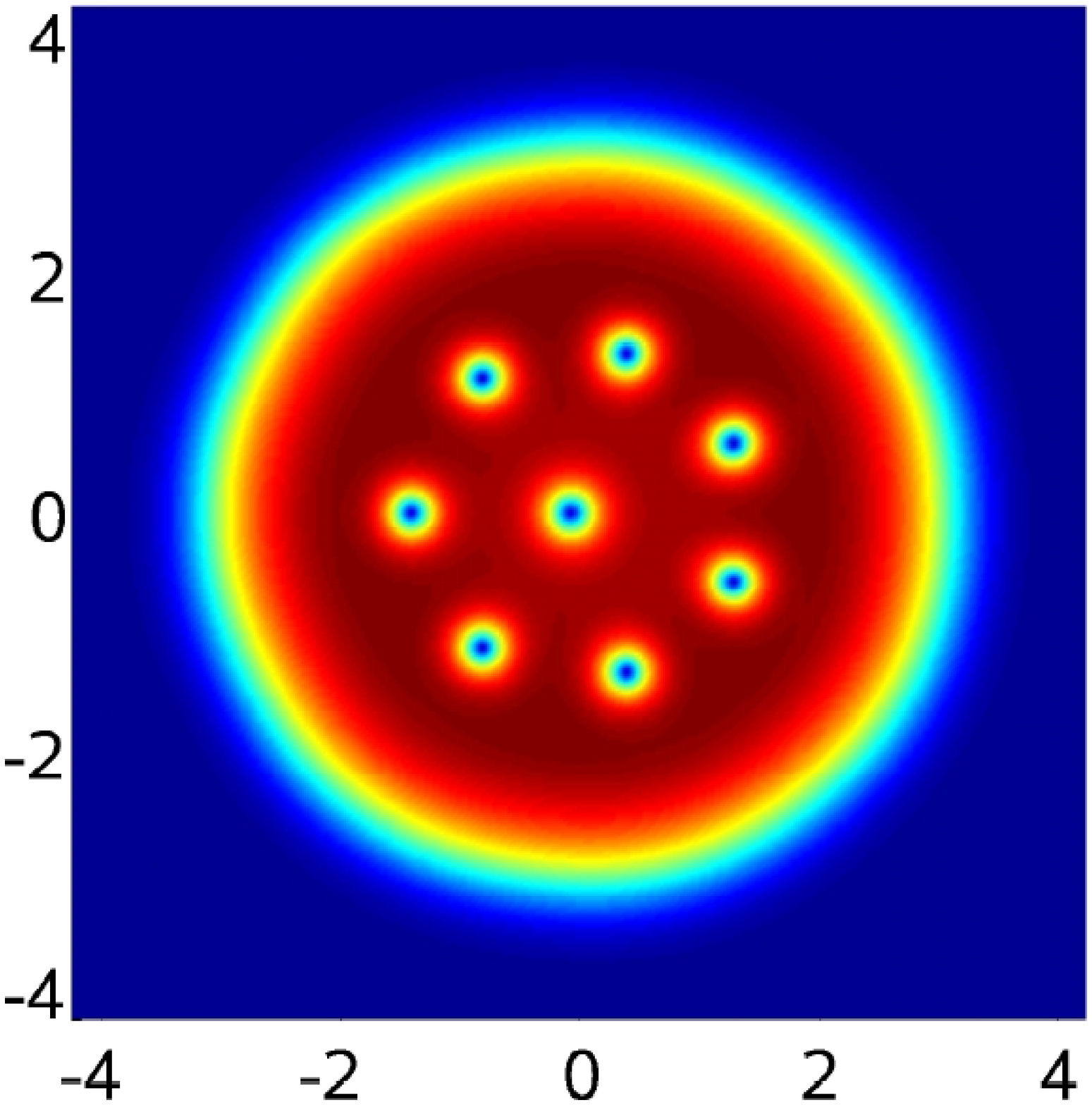}
\caption{Vortex array case. Initial state with 6 vortices and final converged state with an array of 7 vortices. Contours of atomic density $|u|^2$.}
\label{fig-array}
\end{figure}

The harmonic trapping potential physically sets an upper bound for the rotation frequency, since for $\Omega=1$ the centrifugal force balances the trapping force and the confinement of the condensate vanishes. To overcome this limitation, different forms of the trapping potential are currently experimentally and theoretically studied. We use in the third numerical experiment a combined harmonic-plus-quartic potential (see also \cite{kasamatsu1,danaila2,danaila3,fetter}) with the following parameters
\begin{equation}
g = 500, \quad V_{trap}=r^2/2+r^4/4, \quad \Omega = 2.
\end{equation}
The Thomas-Fermi radius is for this case $R_\TF^{\Omega} = 3.40$.
The computational domain is circular of radius $R_{max}= 1.25 R_\TF^{\Omega}$.
The initial state contains a central vortex plus an array of 6 vortices equally distributed on the circle of radius $0.25 R_{max}$. All the vortices have a winding number $m=1$, except the first vortex that has $m=2$ (Fig. \ref{fig-array}). Since vortices with winding number $m>1$ are not physically stable, the $m=2$ vortex will split into two singly quantized vortices. The final state contains therefore a central vortex an array of 7 vortices (Fig. \ref{fig-array}).  The convergence test is relaxed to $\varepsilon=10^{-6}$.

\begin{table}[!h]
\centering
\begin{tabular}{ | l | l | l | l | l | l | } \hline
    gradient&   $N$ &       $n$     &   CPU    &    E(u)    &       $L_z$   \\
\hline \hline
    $H$ &   $2^6$   &   610 &   89.07   &   11.2679 &   6.4549 \\
\hline
    $H_A$   &   $2^6$   &   459 &   79.76   &   11.2670 &   6.4576  \\
\hline \hline
    $H$ &   $2^7$   &   530 &   466.34  &   11.2971 &   6.4603  \\
\hline
    $H_A$   &   $2^7$   &   442 &   447.92  &   11.2959 &   6.4603  \\
\hline \hline
    $H$ &   $2^8$   &   539 &   2.4760e3&   11.2990 &   6.4605      \\
\hline
    $H_A$   &   $2^8$   &   441 &   2.245e3&    11.2977 &   6.4691  \\
\hline

\end{tabular}\\
\caption{Vortex array case.  Algorithm efficiency and characterization ($E(u), L_z(u)$) of the converged state  state  for the finite difference implementation (variable time step computation).}
\label{tab-exp3-FD}
\end{table}

\begin{table}[!h]
\centering
\begin{tabular}{ | l | l | l | l | l | l | } \hline
    gradient&   $M$/triangles   &       $n$     &   CPU    &    E(u)    &       $L_z$   \\
\hline \hline
    $H$ &   100/1762    &   507 &   42.28   &   12.0553 &   6.1297 \\
\hline
    $H_A$   &   100/1762    &   330 &   27.70   &   12.1413 &   6.1654  \\
\hline \hline
    $H$ &   200/7064    &   418 &   138.53  &   11.5341 &   6.3920  \\
\hline
    $H_A$   &   200/7064    &   270 &   90.11   &   11.6171 &   6.4135  \\
\hline \hline
    $H$ &   400/27604   &   420 &   550.10  &   11.4017 &   6.4641  \\
\hline
    $H_A$   &   400/27604   &   262 &   346.87 &    11.4846 &   6.4840  \\
\hline
\end{tabular}
\caption{Vortex array case. Algorithm efficiency and characterization ($E(u), L_z(u)$) of the converged state for the  finite element implementation.
The maximum allowed time step is 0.1 for the $H$ gradient and 0.2 for the $H_A$ gradient.}
\label{tab-exp3-FE}
\end{table}

Tables \ref{tab-exp3-FD} and \ref{tab-exp3-FE} show that the converged state is the same for both finite difference and finite element implementations. The $H_A$ method has better stability properties and allows a CPU time gain up to 36\%. This gain was expected since $\Omega$ is large for this case.

\subsubsection{Giant vortex and Abrikosov vortex lattice}

\begin{figure}[!h]
\centering
\includegraphics[width=0.90\columnwidth]{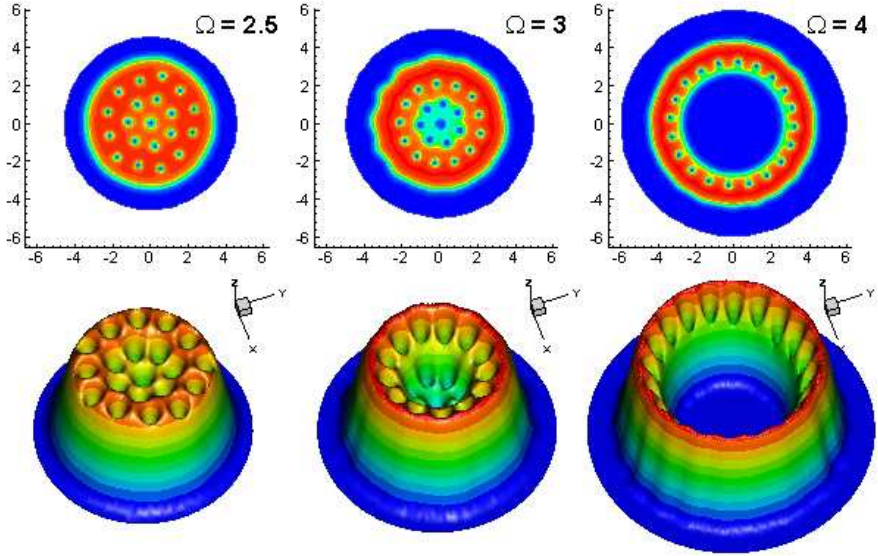}
\caption{Giant vortex case. Converged states for $\Omega=2.5, 3, 4$ showing the formation of a hole in the condensate (giant vortex) for high rotation rates Contours of atomic density $|u|^2$.}
\label{fig-giant}
\end{figure}

Finally, to show that the new method has the capability to handle more complicated cases, we produce the giant vortex using the $H_A$ gradient in conjunction with the new projection method proposed to enforce the mass constraint.  We use the parameters of the previous numerical experiment (harmonic-plus-quartic trapping potential) and progressively increase $\Omega$ from 2 to 4. Each computation starts from an initial field representing the converged state previously obtained for a lower value of $\Omega$.
The transition from  a vortex lattice to the giant vortex is observed (Fig. \ref{fig-giant}).  The giant vortex is a hole in the condensate (the atomic density goes to zero inside) with multiple phase defects.
This particular vortex structure, theoretically analyzed in numerous studies \cite{kasamatsu1,danaila2,danaila3,fetter},
was captured using both the finite elements and finite difference simulations.

\begin{figure}[!h]
\centering
\includegraphics[width=0.82\columnwidth]{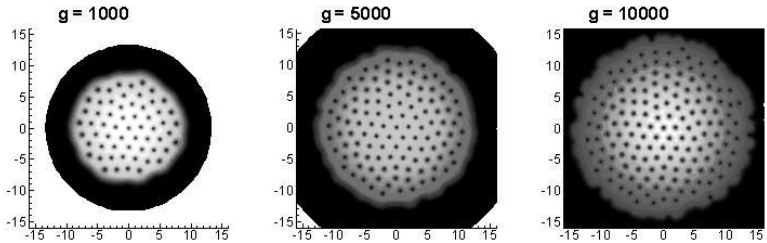}
\caption{Abrikosov  vortex lattice case. Converged states for $\Omega=0.95$ and increasing values of the interactions constant $g$. Finite elements computations using mesh adaptivity. Contours of atomic density $|u|^2$.}
\label{fig-abrikosov}
\end{figure}

A last complex computational case is illustrated in Fig. \ref{fig-abrikosov}.
For a harmonic trapping potential and high rotation frequency ($\Omega=0.95$) an Abrikosov vortex lattice forms in the condensate. The difficulty in computing this case in the strong-interaction regime (large values of $g$) comes from the fact that the condensate becomes larger and the vortex lattice denser when the value of $g$ is increased. In order to increase convergence,
each computation starts from an initial field representing the converged state obtained for a lower value of $g$. During the iterative process, new vortices nucleate at the boundaries and slowly move towards their final equilibrium locations. In computing such configurations, containing several hundreds of vortices, the adaptive mesh refinement capabilities of FreeFem proved very helpful in reducing the computational time and correctly capturing vortex positions.

\section{Summary}

The numerical study of a rotating Bose-Einstein condensate has been the subject of many numerical studies, both in two dimensions (2D) and three dimensions (3D).  Since most of the studies \cite{du,kasamatsu1,kasamatsu2,danaila1,danaila2,danaila3,bao,bao1} use the imaginary time propagation method (equivalent to the gradient flow model \eqref{eq-grad-flow}), there are few studies using direct minimization by
Sobolev gradient methods. Nevertheless, replacing the ordinary $L^2$ gradient in a descent method with the Sobolev $H^1$ gradient  proved effective in minimizing the 3D Gross-Pitaevskii energy  \cite{garcia1,garcia-ripoll} or simpler Schr{\"o}dinger type functionals \cite{lookman}.

In this work we introduced a new inner product ($H_A$) to equip the domain of the GP energy functional with rotation and derived the
corresponding  gradient. We demonstrated that numerical performance is enhanced by replacing in the descent method the $L^2$ or $H^1$ gradients with the gradient obtained from the $H_A$ inner product. The gain in computational time proved very important when configurations with high rotation rates are computed.
We also introduced a new projection method to enforce the mass constraint. This method avoids more complicated approaches using an energy functional with a penalty term, or the traditional normalization method that performs the descent over a path of functions with an imposed norm.

These two new tools allowed to implement robust descent methods using finite difference and finite element spatial discretization. Both numerical settings proved very efficient in computing various complex two-dimensional configurations of rotating Bose-Einstein condensates.

We finally emphasize the fact that the new gradient and  projection method for the mass constraint have a more general interest and could be also used in conjunction with existing numerical schemes (such as sophisticated time stepping procedures) to study the energy minimization of Gross-Pitaevskii type functionals.

\section*{Acknowledgements}

P. Kazemi acknowledges support from CNRS and thanks Professor Sylvia Serfaty for hosting her at Laboratoire Jacques-Louis Lions. She also thanks
Insitut f{\"u}r Quantenphysik, Universit{\"a}t Ulm for their generous hospitality during her visits. I. Danaila acknowledges helpful discussions with Professor F. Hecht in optimizing FreeFem scripts.


\begin{thebibliography}{99}
\bibitem{aboshaerr} J. R. Abo-Shaeer ,   C. Raman, J. M. Vogels, W. Ketterle, Observation of vortex lattices in Bose-Einstein condensates, Science, 292 (2001), pp. 476-479.

\bibitem{adams} R. Adams, \textit{Sobolev Spaces}, Academic, New York, 1975.

\bibitem{danaila1} A. Aftalion and I. Danaila, Three-dimensional vortex conigurations in a rotating Bose Einstein condensate,
Phys. Rev. A, 68 (2003), 023603.

\bibitem{danaila2} A. Aftalion and I. Danaila, Giant vortices in combined harmonic and quartic traps, Phys. Rev. A, 69 (2004), 033608.

\bibitem{du} A. Aftalion and Q. Du, Vortices in a rotating Bose-Einstein condensate: critical
angular velocities and energy diagrams in the Thomas-Fermi regime, Phys. Rev. A, 64 (2001), 063603.

\bibitem{alouges1} F. Alouges, A new algorithm for computing liquid crystal stable configurations: the harmonic mapping case, SIAM J. Numer. Anal., 34 (1997), pp. 1708-1726.

\bibitem{alouges} F. Alouges and C. Audouze, Preconditioned gradient flows for nonlinear eigenvalue problems and applications to the Hartree-Fock functional, Numerical Methods for Partial Differential Equations, 25 (2008), pp. 380-400.

\bibitem{anderson} M. H. Anderson, J. P. Ensher, M. H. Matthews, C. E. Wieman, and E. A. Cornell, Observation of Bose-Einstein condensation in a dilute atomic vapor, Science,  269 (1995), pp. 198-201.

\bibitem{bao} W. Bao and Q. Du, Computing the ground state solution of Bose--Einstein condensates by a normalized gradient flow,
SIAM J. Sci. Comput. 25 , 5 (2004), pp. 1674-1697.

\bibitem{bao2008} W. Bao and J. Shen, A generalized-Laguerre-Hermite pseudospectral method for computing symmetric and central vortex states in Bose-Einstein condensates, Journal of Computational Physics,  227 (2008),pp. 9778-9793.

\bibitem{bao1} W. Bao and W. Tang, Numerical solution of the Gross-Pitaevskii equation for Bose-Einstein condensation,
Journal of Computational Physics, 187 (2003), pp. 230-254.

\bibitem{bradley} C. C. Bradley, A. Sacket, J. J. Tollett, R. G. Hulet, Evidence of Bose-Einstein condensation in an atomic gas with attractive interactions, Phys. Rev. Lett. , 75 (1995), pp. 1687-1690.

\bibitem{danaila3} I. Danaila, Three-dimensional vortex structure of a fast rotating Bose-Einstein condensate with harmonic-plus-quartic confinement, Phys. Rev. A, 72 (2005), 013605.

\bibitem{davis} K. B. Davis, M. O. Mewes, M. R. Andrews, N. J. Druten, D. S. Durfee, D. M. Kurn, and W. Ketterle, Bose-Einstein condensation in a gas of sodium atoms, Phys. Rev. Lett., 75 (1995), pp. 3969-3973.

\bibitem{fetter} A. L. Fetter, Rotating trapped Bose-Einstein condensates, Laser Physics 18, 1 (2008), pp. 1 - 11.

\bibitem{fetter1} A. L. Fetter, B. Jackson, S. Stringari, Rapid rotation of a Bose-Einstein condensate in a harmonic plus quartic trap, Phys. Rev. A, 71 (2005) , 013605.

\bibitem{garcia-ripoll} J. J. Garc\'ia-Ripoll and V. M. P{\'e}rez-Garc\'ia, Optimizing Schr{\"o}dinger Functionals Using Sobolev gradients: application to quantum mechanics and nonlinear optics, SIAM J. Sci. Comput., 23 (2001), pp. 1315-1333.

\bibitem{garcia1} J. J. Garc\'ia-Ripoll and V. M. P{\'e}rez-Garc\'ia, Vortex bending and tightly packed vortex lattices in Bose-Einstein condensates, Phys. Rev. A,  64 (2001), 053611.

\bibitem{haljan} P. C. Haljan  , I. Coddington, P. Engels, E. A. Cornell, Driving Bose-Einstein condensate vorticity with a rotating normal cloud, Phys Rev Lett., 87 (2001), pp. 210403-210407.

\bibitem{freefem} F. Hecht, O. Pironneau, A. Le Hyaric and K. Ohtsuke, FreeFem++ (manual), www.freefem.org.

\bibitem{kasamatsu1} K. Kasamatsu, M. Tsubota, M.  Ueda, Giant hole and Circular superflow in a fast rotating BEC,
Phys. Rev. A, 66 (2002), 053606.

\bibitem{kasamatsu2} K. Kasamatsu, M. Tsubota, M. Ueda, Vortex lattice formation in a rotating Bose-einstein condensate,
Phys. Rev. A, 65 (2002), 023603.

\bibitem{pkme} P. Kazemi and M. Eckart, \textit{Minimizing the Gross-Pitaevskii energy functional with the Sobolev gradient -- Analytical and numerical results}, to appear in the International Journal of Computational Methods, preprint available arXiv:0906.3206.

\bibitem{lieb} E. H. Lieb and R. Seiringer, Derivation of the Gross-Pitaevskii equation for rotating Bose gases,
Comm. in Math. Physics, 264 , 2 (2006), pp. 505-537.

\bibitem{madison} K. W. Madison, F. Chevy, W. Wohleben, J. Daribard, Vortex formation in a stirred Bose-Einstein condensate, Phys. Rev. Lett,. 84 (2000), pp. 806-809.

\bibitem{matthews} M. R. Matthews, B. P. Anderson, P. C. Hajlan, D. S. Hall, M. J. Holland, J. E. Williams, C. E. Weiman, E. A. Cornell, Watching a superfluid untwist itself: recurrence of rabi oscillations in a Bose-Einstein condensate, Phys. Rev. Lett., 83 (1999), pp. 3358-3361.


\bibitem{jwn} J. W. Neuberger, Sobolev Gradients and Differential Equations, Lecture notes in mathematics, 1670,  Springer, Berlin / Heidelberg, 1997.

\bibitem{pierre} M. Pierre, Newton and conjugate gradient for harmonic maps from the disc into the sphere, ESAIM: Control, Optimisation and Calculus of Variations, 10 (2004), pp. 142-167.

\bibitem{lookman} N. Raza, S. Sial , S. S. Siddiqi , T. Lookman, Energy minimization related to the nonlinear Schr{\"o}dinger equation, Journal of Computational Physics, 228 (2009), pp. 2572-2577.

\end{thebibliography}
\end{document}